# Accurate Prostate Cancer Detection and Segmentation on Biparametric MRI using Non-local Mask R-CNN with Histopathological Ground Truth

Zhenzhen Dai, Ivan Jambor, Pekka Taimen, Milan Pantelic, Mohamed Elshaikh, Craig Rogers, Otto Ettala, Peter Boström, Hannu Aronen, Harri Merisaari, Ning Wen

zdai1@hfhs.org, ivjamb@utu.fi, pepeta@utu.fi, milan@rad.hfh.edu, melshai1@hfhs.org, crogers2@hfhs.org,

otto.ettala@tyks.fi, peter.bostron@tyks.fi, hannu.aronen@utu.fi, haanme@utu.fi, nwen1@hfhs.org

**Purpose:** We aimed to develop deep machine learning (DL) models to improve the detection and segmentation of intraprostatic lesions (IL) on bp-MRI by using whole amount prostatectomy specimen-based delineations. We also aimed to investigate whether transfer learning and self-training would improve results with small amount labelled data.

**Methods:** 158 patients had suspicious lesions delineated on MRI based on bp-MRI, 64 patients had ILs delineated on MRI based on whole mount prostatectomy specimen sections, 40 patients were unlabelled. A non-local Mask R-CNN was proposed to improve the segmentation accuracy. Transfer learning was investigated by fine-tuning a model trained using MRI-based delineations with prostatectomy-based delineations. Two label selection strategies were investigated in self-training. The performance of models was evaluated by 3D detection rate, dice similarity coefficient (DSC), 95 percentile Hausdrauff (95 HD, mm) and true positive ratio (TPR).

**Results**: With prostatectomy-based delineations, the non-local Mask R-CNN with fine-tuning and self-training significantly improved all evaluation metrics. For the model with the highest detection rate and DSC, 80.5% (33/41) of lesions in all Gleason Grade Groups (GGG) were detected with DSC of 0.548±0.165, 95 HD of 5.72±3.17 and TPR of 0.613±0.193. Among them, 94.7% (18/19) of lesions with GGG > 2 were detected with DSC of 0.604±0.135, 95 HD of 6.26±3.44 and TPR of 0.580±0.190.

**Conclusion:** DL models can achieve high prostate cancer detection and segmentation accuracy on bp-MRI based on annotations from histologic images. To further improve the performance, more data with annotations of both MRI and whole amount prostatectomy specimens are required.

## 1. Introduction

Prostate cancer (PCa) accounts for more than 20% of new cancer diagnoses in men, with an estimated 191,930 new cases and 33,330 deaths expected in 2020 (1). Automatic detection and segmentation of intraprostatic lesions (IL) on MRI can optimize radiology workflow (2) and provide radiologists with preoperative knowledge for diagnosis, risk stratification, staging and appropriate therapeutic treatment guidance, including focal therapy with cryoablation or laser ablation, or simultaneous integrated boost treatment to enhance the radiation dose to ILs (3, 4).

Three major factors were considered in the implementation of deep machine learning (DL) models to accurately detect and segment ILs. First, MRI has been found to mis-classify and consistently underestimate the size and extent of PCa (5). Histopathological tissue analysis is considered as the gold standard for cancer diagnosis and determination of the malignancy level on a microscopic level (6). It offers significant benefits for automatic detection and segmentation of ILs, if ground truth delineations are obtained based the whole amount prostatectomy specimen. Second, the convolutional operations used in most DL models processes within a local neighbourhood of the input, and the localization and segmentation accuracy are discounted due to imperfect alignment of different imaging modalities. Pooling and resizing operations often cause information loss in segmentation models and the lost information normally contains the opaque regions between a lesion and surrounding normal tissues, leading to inaccurate defining of the boundary. Lastly, it is extremely time-consuming and challenging to register the histology slices to MRIs and label them in a multidisciplinary collaboration. Therefore, there has been no progress yet on DL based segmentation using cross modality image domains due to the small volume of annotated data.

In this work, we improved the detection and segmentation of PCa on bp-MRI by using delineations based on histopathological images as ground truth. A non-local Mask R-CNN was proposed to improve the segmentation accuracy by addressing imperfect image registration. We also investigated a variety of techniques including fine-tuning and self-training which are often used to alleviate the small data problem to determine whether they would help improve the model in our experimental settings. Because it is arguable whether semi-supervised learning and whether selecting unlabelled data with efforts and high confidence is superior, we also experimented with two label selection strategies and investigated how selection properties impacted performance.

## 2 Material and Method

### 2.1 Patients Cohorts

DL models were obtained by using delineations based on two different imaging modalities, bp-MRI (Cohort 1) and whole mount prostatectomy specimen (Cohort 2), as ground truth. 158 Cohort 1 patients from two datasets had suspicious ILs delineated on bp-MRIs based on the reading from bp-MRIs only. Cohort 1a included 97 patients from the SPIE-AAPM-NCI Prostate MR Gleason Grade Group Challenge (PROSTATEx-2 Challenge) (7-9). Cohort 1b included 61 patients from the IMPROD Trial (10) who underwent bp-MRI only. Cohort 2 included a separate 64 patients from (10) who underwent prostatectomy with whole-mount prostatectomy section following bp-MRI and biopsy. 40 (Cohort 3) private patients were unlabelled. Table 1 summarizes the patient demographics and clinical characteristics.

**Cohort 1a: bp-MRI Based Delineations**
Suspicious ILs were delineated on T2-weighted images (T2WI) by an experienced radiologist based on the interpretation of T2WI and apparent diffusion coefficient (ADC) map respectively and were then aggregated, without knowledge of biopsy or other clinical findings such as PSA. Biopsy results were provided by the dataset but were independent from the suspicious ILs delineations.

**Cohort 1b: bp-MRI Based Delineations**
Suspicious ILs were delineated by an experienced radiologist (12 years of prostate MRI experience) using integrated information from bp-MRIs without reference to biopsy or other laboratory parameters. Biopsy result for each bp-MRI based delineation was provided by the dataset (10).

**Cohort 2: Whole Amount Prostatectomy Specimen Based Delineation**
ILs of Cohort 2 were identified and delineated on T2WIs by using whole-mount prostatectomy specimen as reference standard. The prostate glands were fixed in 10% buffered formalin for 24–48 hours after robotic radical histology. Prostate surfaces were inked with different colors to preserve the orientation of the prostate gland which allows for the correlation with MRI datasets. Whole-mount histology macro-sections were obtained at 4–6 mm intervals transversely in a plane perpendicular to the long axis of the prostate gland in the superior–inferior direction. Four μm whole-mount sections from each macro-block were stained with hematoxylin and eosin (HE) (10, 11). The HE-stained histological slides were first reviewed by one staff board-certified pathologist and then re-reviewed by an experienced genitourinary pathologist. If there were controversy between the two reviews, the slices were reviewed by a third genitourinary pathologist and consensus was reached between the genitourinary pathologists. Whole-mount histology sections were matched manually with T2WIs based on anatomical landmarks such as benign prostatic nodules and urethra. Delineations were subsequently corrected on T2WIs using the whole-mount histology as reference standard. When performing delineations using whole-mount prostatectomy as the reference standard, the research fellow working in conjunction with the pathologist did not have access to prospective lesion delineation, to limit possible recall bias. If a lesion was discontinued and the areas were within 1 mm of each other, the areas were considered as a conglomerate lesion and treated as one lesion.

**Table 1.** Demographics and clinical characteristics of 262 patients from three cohorts.

| MRI-based Cohort 1 | | prostatectomy-based Cohort 2 | Unlabeled Cohort 3 |
|---|---|---|---|
| Cohort 1a | Cohort 1b | | |
| Gleason Grade Group, n (%)<br>1  21 (21.6%)<br>2  30 (30.9%)<br>3  17 (17.5%)<br>4  5 (5.2%)<br>5  6 (6.2%)<br>NA  18 (18.6%) | Gleason Grade Group, n (%)<br>1  13 (17.3%)<br>2  13 (17.3%)<br>3  7 (9.3%)<br>4  5 (6.7%)<br>5  1 (1.3%)<br>Benign  36 (48.0%) | Gleason Grade Group, n (%)<br>1  1 (1.6%)<br>1+  2 (3.1%)<br>2  45 (43.7%)<br>2+  2 (3.1%)<br>3  15 (20.3%)<br>3+  8 (12.5%)<br>4  14 (4.6%)<br>5  1 (10.9%) | Gleason Grade Group, n (%)<br>NA |
| MRI scanner<br>MAGNETOM Trio and Skyra | MRI scanner<br>Verio, Siemens, Erlangen, Germany | MRI scanner<br>Verio, Siemens, Erlangen, Germany | MRI scanner<br>Ingenia; Philips Medical System, Best, the Netherlands |
| T2WI<br>Voxel Resolution:<br>$0.5 \times 0.5 \times 3.6$ mm³<br><br>ADC<br>b-values:<br>50, 400, 800 s/mm²<br>Voxel Resolution:<br>$2 \times 2 \times 3.6$ mm³ | T2WI<br>Voxel Resolution:<br>$0.625 \times 0.625 \times 3$ mm³<br><br>ADC<br>b-values:<br>0, 100, 200, 300, 500 s/mm²<br>Voxel Resolution:<br>$2.0 \times 2.0 \times 3.0$ mm³ | T2WI<br>Voxel Resolution:<br>$0.625 \times 0.625 \times 3$ mm³<br><br>ADC<br>b-values:<br>0, 100, 200, 300, 500 s/mm²<br>Voxel Resolution:<br>$2.0 \times 2.0 \times 3.0$ mm³ | T2WI<br>Voxel Resolution: $0.42 \times 0.42 \times 2.4$ mm³<br><br>ADC<br>b-values:<br>0, 1000 s/mm²<br>Voxel Resolution:<br>$1.79 \times 1.79 \times 0.56$ mm³ |

## 2.2. Image Pre-processing

The N4 bias field correction (12) was applied to both T2WI and ADC map to correct the low frequency intensity inhomogeneity. ADC maps were registered to T2WI by image transformation. For Cohort 1a and Cohort 3 patients and unlabeled patients, rigid transformation was used (MIM software, version 6.7) and for Cohort 1b and Cohort 2 patients, nonreflective similarity transformation was used (MATLAB, MathWorks). Prostate patches were cropped out from the co-registered T2WI and ADC map given 1 pixel (px) margin from the prostate contour and were combined as the two-channel input of the model. T2WI and ADC map were normalized by subtracting the mean and divided by the standard deviation, histogram equalization was applied to both T2WI and ADC map to enhance the contrast. Finally, the input was resized to the resolution of $256 \times 256 \times 2$ px and the aspect ratio of prostate were preserved by padding with zero.

## 2.3. Post processing - 3D Prediction Aggregation

The model made predictions on each prostate slice from the patient. Each prediction on the slice consisted of a binary mask and a corresponding mask score, which were calculated by the segmentation and prediction branch respectively. The binary mask was model's prediction of lesion, the corresponding mask score was the probability estimate of the binary mask as a lesion. The set of 2D predictions was aggregated into 3D predictions according to their mask scores and spatial connectivity between masks for the patient. First, any 2D predictions with mask scores smaller than 0.7 ($\beta$) were discarded. The rest on each slice were aggregated if the DSC between any two masks was greater than 0.7 ($\gamma$), the two masks were combined, and their mask scores were averaged. After this, 2D predictions were further aggregated across the axial direction and were aggregated into one 3D prediction if, on any two adjacent slices, DSC of the two masks was greater than 0.35 ($\alpha$). The thresholds $\alpha = 0.35$, $\beta = 0.7$, $\gamma = 0.7$ were defined based on our empirical knowledge. The 3D prediction was scored as the maximum mask score inside the aggregated 2D predictions. To increase the detection rate, 3D predictions with top five scores were selected as the final 3D predictions for each patient (top-5 3D predictions).

## 2.4. Non-local and baseline Mask R-CNN

Convolutional operation, which is the essential component in Mask R-CNN and most deep segmentation networks, calculates the response of an input within a local neighborhood, for example, a kernel size of 3, and is consequently sensitive to misalignment of pixels when the receptive field is small. The receptive field becomes larger at deeper layers, but information is lost due to operations such as pooling and resizing. Thus, though the advantage of using multi-parametric MRI in prostate cancer localization and segmentation has been demonstrated (13, 14), the misalignment of T2WI and ADC compromises localization and segmentation accuracy and improvements can be done over the Mask R-CNN baseline. The image registration helps reduce but not eliminate the inherent alignment errors on the voxel level. The non-local network (Figure 1B) described in (15) provided the basis of our non-local Mask R-CNN (Figure 1). It aimed at modelling long-range dependencies across spatial regions of input by computing the response at a position based on the relationships between it and all positions in the input. As shown in Figure 1C, an adapted non-local network shown in Figure 1B was added after the registered T2 and ADC map. To reduce computation cost, the input was down sampled to I: $64 \times 64 \times 2$. The new input $I' = I + f(softmax(\theta(I_i)^T \varphi(I_j))g(I))$, where $f, \theta, \varphi$ and $g$ were embedding functions and were implemented as $1 \times 1$ convolution, were up-sampled to its original size of $256 \times 256 \times 2$. The softmax operation made the non-local network specially related to the self-attention module discussed in (16). In our implementation, softmax was added along both dimension i and j, which was different from (15) where softmax was added along dimension j. Therefore, when synthesizing the ith position, the model was able to consider not only the relationships between the ith position and other positions but also the relationships between all other position pairs. In addition, the attended input $A$ was further concatenated with resizing to each layer of feature pyramids to enrich features and guide precise prediction.

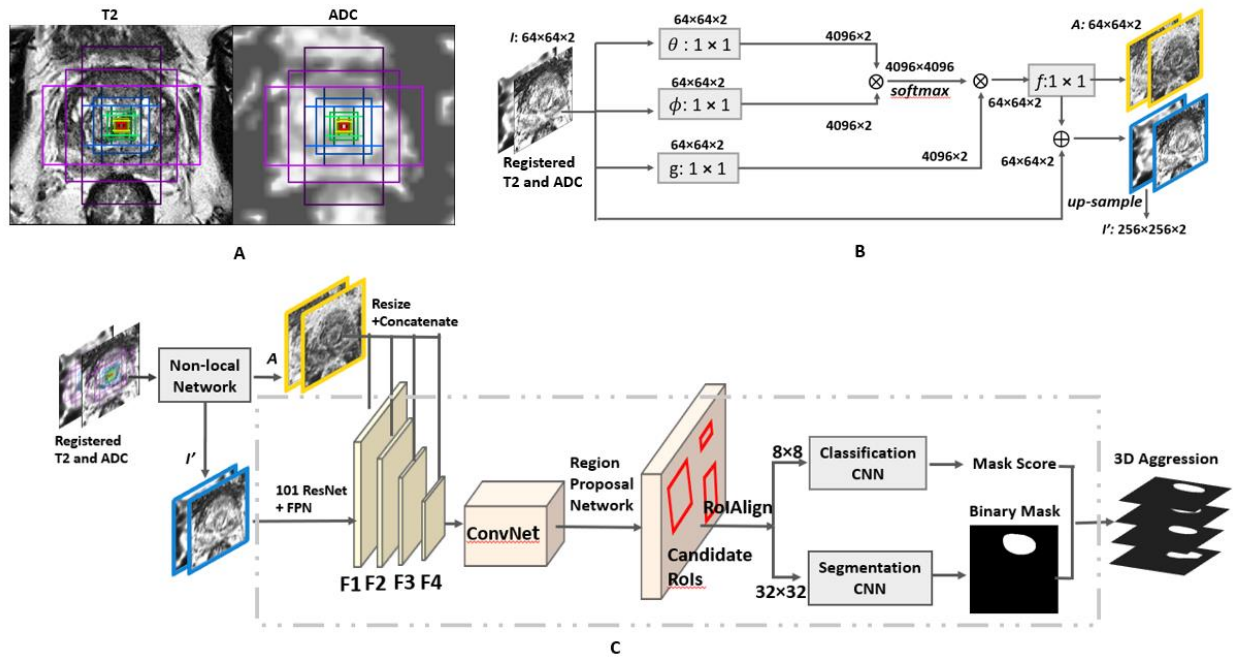

**Figure 1.** (A) The registered T2WI and ADC map with prostate patches cropped out as input. Alignment of the two modalities was still not perfect after registration. (B) Non-local network. It outputs attended input $A$ and new network input $I'$ (C) Architecture of Mask R-CNN baseline is shown in the grey dashed box. 101 layer deep residual network (ResNet) was used as the bottom-up pathway for feature pyramid network (FPN) to extract pyramid feature maps (F1: $64 \times 64 \times 256$, F2: $32 \times 32 \times 256$, F3: $16 \times 16 \times 256$, F4: $8 \times 8 \times 256$) over the new input $I'$. A set of anchors were predefined at each input position and at different scales and each anchor bound to the feature map with corresponding positions and scales. Anchors at the centre point were shown in (A) as boxes of different colors. The pyramid feature maps concatenated with the attended input $A$ were passed to a shared convolution network (ConvNet) to predict the class (background or foreground) and bounding box regression parameters. The region proposal network selected top positive anchors and applied bounding box regression parameters to the selected anchor to generate candidate region of interest (RoI). The RoIAlign layer mapped the candidate RoIs on feature pyramids according to its size and location. Mapped feature maps of each candidate RoI were resized to $8 \times 8$ pixels and $32 \times 32$ pixels and then were passed into the classification and segmentation branch, respectively. The segmentation branch only took positive RoIs selected by the classification branch and generated binary masks which were aggregated into 3D predictions.

### 2.5. Self-training - Selection of Unlabeled Data to Improve Prostatectomy-based Model
We explored the possibility of using unlabeled data to improve the prostatectomy-based model by self-training under two label selection strategies when the amount of labelled data was limited. All models incorporated self-training were fine-tuned from the non-local MRI-based Mask R-CNN (R2).

Data distillation was used to generate new training samples with refined delineations by ensembling the results from a single model run on different transformations (flipping, resizing, rotation, etc.) of unlabeled image sets (17). Such transformations are usually used as data augmentation options in training and are proved to improve single model's accuracy. For a trained model, the best transformation options were selected by detection rate and DSC of the top-5 predictions from fliplr, flipud and two rotation angles on the prostatectomy-based validation patients. The rotation angle was in the range of (0°, 360°) with a step size of 5°. With best transformation options selected, we applied data distillation to each slice and obtained the refined delineation.

Moreover, false positives and false negatives in slices or pixels were depreciated in generating new IL delineations on unlabeled data. On the one hand, the predicted mask scores could be used as a proxy to select predictions above a score threshold in increasing selection confidence. And averaging mask scores from different transformations help remove single prediction bias. On the other hand, additional information from other clinical reference such as biopsy information and MRI-based delineations could be used to select the most confident labels instead of relying only on the prediction by the base model.

The self-training consisted of 4 steps, 1) predicting multiple best transformations of unlabeled data by the trained model; 2) ensembling predictions from multiple transformations. We experimented with 2D combination and 2D voting; 3) selecting ensembled labels as new training labels using certain selection strategy; 4) retraining the model on the union set of the original labelled data and automatically labelled data. In our experiment, R5 was used as the start point to predict new training samples, and the retrained model was used as new start point and iterated the above 4 steps multiple times to investigate how selection properties impacted performance by each strategy after each iteration. We used all unlabeled data in each iteration unless no predictions met the selection criteria and iterated three times. And to combine semi-supervised learning with fine-tuning, each retrained model was obtained by fine-tuning R2.

The two selection strategies were investigated. S1: 2D combination of multiple transformations (Figure S1-B) was applied to each slice followed by the plain 3D aggregation discussed in Section 2.3. Among the top-5 predictions, only the most confident predictions were selected. The criteria were based on biopsy results and its corresponding MRI-based delineation, any Gleason Grade Group (GGG) were considered. 92 patients from the cohort 1 were found having MRI-based delineations as biopsy-proven prostate cancer (PCa) and were reused as unlabeled patients. Only predictions having DSC larger than 0.5 with biopsy-proven MRI-based delineations among the top-5 predictions were selected as new training samples. It should be noted that the approximate location of selected unlabeled data could match with the IL by using biopsy and MRI-based delineations, however, the boundary was defined by the model itself. S2: 2D voting (at least half of predictions from the original and transformed images should agree) of multiple transformations (Figure S1-C) followed by the plain 3D aggregation discussed in Section 2.3. MRI-based validation and testing patients as well as Cohort 3 were used as unlabeled data and totally 94 patients were used. Therefore, equivalent number of patients were used for both strategies. Compared S1 and S2, S1 minimized the number false positives yet it might wrongly reject some true positives as MRI underestimates IL. S2 keeps all top-5 3D predictions by the model to maximize the number of true positives yet large number of false positives were also introduced in selected labels. And 2D voting was used to reduce false positives to some extent. S1 was designed to investigate whether keeping the most confident labels could resulted in better performance than S2.

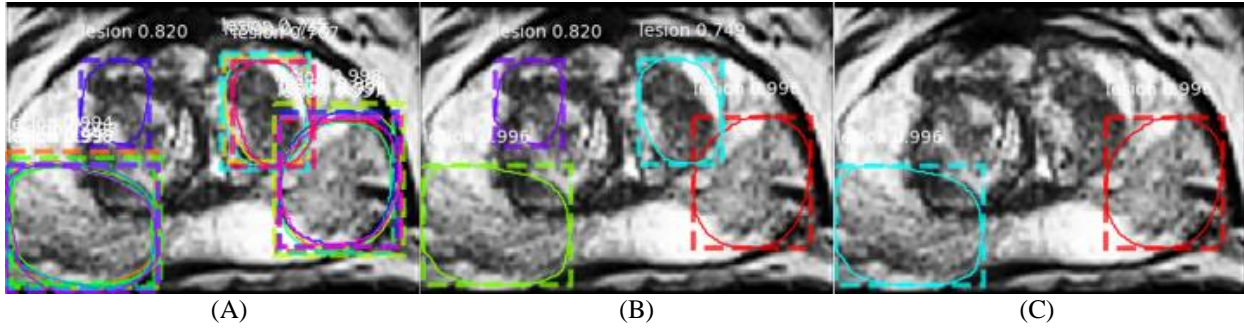

(A)　　　　　　　　　　　　(B)　　　　　　　　　　　　(C)

**Figure S1**. Ensembling methods of multiple transformations. (A) Predictions on original and transformed images. Predictions on transformed images were transformed back to the original coordinates to display. (B) 2D Combination. All predictions from original and transformed images were kept. Predictions were ensembled if DSC between each other was greater than 0.7. Masks and scores were averaged to ensemble. A cut-off threshold of 0.5 re-transferred the mask into binary. (C) 2D Voting. The final predictions were voted by the predictions on the original and transformed images. Suppose N (N = 5 in this plot) was the number of original and transformed images, if more than N/2 predictions agreed, the predictions were kept and ensembled. Predictions were regarded as agreed if DSC between each other was larger than 0.7. The masks and scores inside the agreed predictions were averaged. A cut-off threshold of 0.5 re-transferred the averaged mask into binary while a score cut-off threshold of 0.85 removed predictions with low scores. On (A), in the peripheral zone, 4 predictions agreed on the left side and five on the right side, and the average scores were above 0.85 thus both were kept in (C). In the central gland, there was only one prediction on the left lateral side and was therefore rejected. There were 3 predictions agreed on the right lateral side, but the average score was below 0.85 thus the prediction was rejected.

## 2.6. Application of deep machine learning network

The DL network took co-registered T2-weighted imaging (T2WI) and apparent diffusion coefficient (ADC) map with prostate patches cropped out as input (Section 2.2. and Figure 1A) and output top-5 3D predictions of lesions (Section 2.3. and Figure 1C). Three 3D aggregation thresholds ($\alpha$, $\beta$, $\gamma$) were used to control the output predictions (Section 2.3.). Our development of DL networks was built upon (18). Adam optimizer was used with learning rate initiated to 0.0001, $\beta_1$ to 0.9 and $\beta_2$ to 0.999. Augmentation options included flipping left-right (fliplr), flipping up-down (flipud), rotation (a random angle from -80° to 80°) and translation in either vertical direction or horizontal direction (random pixels from 0 to 20). All augmentation options were randomly applied with the probability of 0.5. Models were trained using a batch size of one.

The training-validation-testing approach was used throughout our experiment. Each model was trained with 150 epochs and each epoch included iterations of number of slices in corresponding training set. Model parameters were saved each epoch. Model selection and evaluation were performed after the 3D prediction aggregation (Section 2.3.). The best model was selected by 3D detection rate (primary criteria) and 3D dice similarity coefficient (DSC) (secondary criteria) of the top-5 3D predictions on the validation set and were evaluated by detection rate, 3D DSC, 3D 95th percentile hausdorff distance (95 HD) and 3D true positive ratio (TPR) on the testing set. The prostate patch was resized back to its corresponding original resolution before calculating DSC, 95 HD (mm) and TPR.

Both non-local Mask R-CNN and baseline Mask R-CNN were trained from scratch in a fully-supervised fashion using the MRI-based training patients, which resulted in MRI-based baseline Mask R-CNN (R1) and MRI-based non-local Mask R-CNN (R2); the prostatectomy-based training patients resulted in prostatectomy-based baseline Mask R-CNN (R3) and prostatectomy-based non-local Mask R-CNN (R4).

In order to investigate transfer learning and how it impacted performance, R2 was fine-tuned with prostatectomy-based training patients, resulting in R5. Next, two label selection strategies S1, S2 (Section 2.5.) were investigated with self-training (Section 2.5.). Each strategy was tested iteratively; the model trained in the current iteration was used to predict the unlabeled data with the same selection strategy for the next iteration. Each selection strategy was iterated three times resulting in S1-1 ~ S1-3 and S2-1 ~ S2-3, respectively, to evaluate the cumulative effect on model performance.

Models obtained using delineations based on different imaging modalities as ground truth and training techniques are summarized in Table 2. Evaluation metrics are shown in Table 3. Specificity analysis was not included in this work, as top-5 predictions were used to maximize detection rate and limit the number of false positives.

One tailed statistical analysis was performed to compare results from different models. DSC and TPR were compared using t-test; 95 HD was compared using Mann-Whitney test and detection rate was compared using McNemar test. Comparison was made using ILs of all Gleason Grade Group (GGG) and GGG > 2.

**Table 2.** Summary of patients partitioning of training, validation and testing set with corresponding models obtained

| Delineation modality | MRI-based | Prostatectomy-based |
|---|---|---|
| Description | Delineations of suspicious ILs were done using information from bp-MRIs only | IL were identified on prostatectomy specimens and delineations of IL were corrected on T2WIs using the whole-mount histology as reference standard |
| Training patients | 64 Cohort 1a and 40 Cohort 1b patients, 114 suspicious ILs in total | 20 Cohort 2 patients, 42 IL in total |
| Validation patients | 10 Cohort 1a and 5 Cohort 1b patients, 15 suspicious ILs in total | 12 Cohort 2 patients, 16 IL in total |
| Testing patients | 23 Cohort 1a and 16 Cohort 1b patients, 42 suspicious ILs in total | 32 Cohort 2 patients, 41 IL in total. |
| Model obtained | R1 – MRI based baseline Mask R-CNN<br>R2 – MRI based non-local Mask R-CNN<br>*A1 – R1 tested on the 32 prostatectomy-based testing patients<br>*A2 – R2 tested on the 32 prostatectomy-based testing patients | R3 – prostatectomy-based baseline Mask R-CNN<br>R4 – prostatectomy-based non-local Mask R-CNN<br>R5 – R2 fine-tuned with the 20 prostatectomy-based patients<br>S1-1, S1-2, S1-3 – self-training with label selection strategy S1<br>S2-1, S2-2, S2-3 – self-training with label selection strategy S2 |

**Table 3.** Detection rate, DSC, 95 HD and TPR were used as evaluation metrics and were calculated in 3D space. Detection rate is the degree to which a model's predictions concur with ground truth, where *dL* denotes the number of DL model detected lesions (true positive predictions); we considered the detected lesion positive when the ground truth delineation had DSC greater than 0.15 with one of the top-5 3D predictions. If multiple predictions met the criteria, the one with the highest DSC was evaluated. *cL* denotes the number of ground truth delineations. For DL model detected lesions (true positive predictions), DSC, 95[th] percentile HD (95 HD) and TPR were calculated. DSC evaluates how well two binary sets match, where $y_{true}$ is the ground truth delineation, and $\bar{y}_{pred}$ is the model's prediction. A DSC of 1 represents a perfect match of prediction and ground truth. HD measures how far apart subsets within a metric space are from one another, where *a* and *b* are points of sets *A* and *B*, respectively, and $D(a,b)$ is the Euclidean distance between these points. The 95[th] percentile HD indicates that 95% of $\min_{b \in B}(D(a,b))$ is below this cut-off, where A is the ground truth delineation, and B is the prediction by model. TPR measures the proportion of lesions that are correctly identified, where *true positive* is the overlap proportion of prediction by model and ground truth delineation, *positive* is the ground truth delineation.

| Evaluation Metric | Equation |
|---|---|
| $Detection\ Rate$ | $\dfrac{dL}{cL}$ |
| $DSC$ | $\dfrac{2 \cdot <y_{true}, \bar{y}_{pred}>}{<y_{true}, y_{true}> + <\bar{y}_{pred}, \bar{y}_{pred}>}$ |
| $HD(A,B)$ | $HD(A,B) = \max_{a \in A}\left\{\min_{b \in B}(D(a,b))\right\}$ |
| $TPR$ | $\dfrac{true\ positive}{positive}$ |

## 3. Results

Overall, 11 models were trained and investigated for the IL detection and segmentation task. All results were reported on the same 32 Cohort 2 testing patients, 41 IL in total. Among these 41 IL, the number of lesions in GGG 1, 1+, 2, 2+, 3, 3+, and 4 were 4, 2, 16, 1, 7, 3, and 8 respectively. R3 and R4 were reported with adjusted 3D aggregation thresholds with detection rate maximized. Key results were summarized here. Considering detection rate and DSC, the highest performance was achieved by S1-1 for all GGG, with detection rate of 80.5% (33/41), DSC of 0.548 ± 0.165, 95 HD (mm) of 5.72 ± 3.17 and TPR of 0.613 ± 0.193. Among them, 94.7% (18/19) lesions with GGG > 2 were detected with DSC of 0.604 ± 0.135, 95 HD (mm) of 6.36 ± 3.44 and TPR of 0.580 ± 0.190. Considering detection rate and TPR, the highest performance was achieved by S2-3 with detection rate of 80.5% (33/41), DSC of 0.513 ± 0.191, 95 HD (mm) of 5.81 ± 2.89 and TPR of 0.749 ± 0.190. Among them, 84.2% (16/19) lesions with GGG > 2 were detected with DSC of 0.631 ± 0.122, 95 HD (mm) of 6.16 ± 3.08 and TPR of 0.746 ± 0.165.

**Performance using different 3D aggregation thresholds**
We compared non-local Mask R-CNN performance with respect to different 3D aggregation thresholds on R4, R5, and label selection strategies S1-3 and S2-3 respectively. As shown in Figure 2, $\alpha = 0.35$ yielded the best detection and segmentation performance, all GGG included. With the empirical thresholds ($\alpha = 0.35$, $\beta = 0.7$, $\gamma = 0.7$), fine-tuning (R5) led to an improvement over the non-local Mask R-CNN trained from scratch (R4), with detection rate from 65.9% to 75.6% (p = 0.171), DSC from 0.503 to 0.543 (p = 0.183), 95 HD (mm) from 6.91 to 6.28 (p = 0.323) and TPR from 0.597 to 0.625 (p = 0.306) Adjusting thresholds to $\alpha = 0.35$, $\beta = 0.3$ and $\gamma = 0.3$ specially led to an increased detection rate of R4, from 65.9% to 75.9% (p = ***0.037***), however, it didn't yield any significant improvement in other evaluation metrics. The evaluation metrics were most affected by $\alpha$, while the model from scratch was also sensitive to $\gamma$. When increasing $\alpha$, there was a trade-off between detection rate and DSC. However, if $\alpha$ was too high (e.g. increase $\alpha$ to 0.75), both metrics were decreased.

**Comparison of MRI-based and prostatectomy-based models**

When prostatectomy-based delineations were used as ground truth in the training, TPR was significantly improved for both GGG and GGG > 2. For all GGG, when using prostatectomy-based ground truth, baseline Mask R-CNN improved detection rate (R3 compared with A1) from 56.1% to 68.3% (p = 0.114), DSC from 0.400 to 0.429 (p = 0.259), 95 HD (mm) from 8.70 to 7.65 (p = 0.130) and TPR from 0.311 to 0.558 (p = ***1.42E-04***), as shown in Table 4 (c1); non-local Mask R-CNN improved detection rate (R4 compared with A2) from 63.4% to 78.0% (p = 0.057), DSC from 0.427 to 0.504 (p = ***0.033***), 95 HD (mm) from 7.93 to 6.66 (p = ***0.032***) and TPR from 0.372 to 0.589 (p = ***9.64E-05***), as shown in Table 4 (c2). For all GGG (Table 4), all prostatectomy-based non-local models (R4, R5, S1-1 and S2-3) showed significant improvement over the MRI-based baseline Mask R-CNN (A1) in all evaluation metrics. For both all GGG (Table 4) and GGG > 2 (Table 5), S1-1 and S2-3 significantly improved the MRI based models (both baseline and non-local Mask R-CNN, A1 and A2) in all evaluation metrics with only one exception of detection rate of S2-3.

**Performance of prostatectomy-based models with non-local Mask R-CNN and training techniques**
The baseline prostatectomy-based Mask R-CNN (R3) didn't show a significant improvement over MRI-based models, except for TPR. For prostatectomy-based models and for all GGG (Figure 3), non-local Mask R-CNN (R4) and fine-tuned non-local Mask R-CNN (R5) significantly improved baseline Mask R-CNN (R3) in terms of DSC, from 0.429 to 0.504 (p = ***0.043***) and to 0.543 (p = ***0.005***), respectively. Both two label selection strategies significantly improved R3 in terms of DSC (* in Figure 3B) in all iterations. Label selection strategy S2 significantly improved R3 in terms of TPR (* in Figure 3D), and S2-3 increase TPR of R3 from 0.558 to 0.749 (p = **0.001**). 95 HD (mm) was significantly improved in the first two iterations of S1 and the first iteration of S2. For GGG > 2, DSC of R3 was significantly improved by fine-tuning and both two label selection strategies, from 0.469 to 0.579 (p = ***0.033***) by R5, 0.606 (p = ***0.008***) by S1-1 and 0.631 (p = ***0.002***) by S2-3, respectively. TPR of R3 was significantly improved by label selection strategy S2, from 0.539 to 0.746 (p = ***0.008***) by S2-3.

**Table 4.** Results of models in detecting and segmenting ILs, all GGG. R3, R4 reported with adjusted 3D aggregation thresholds with detection rate maximized. A1, MRI-based baseline Mask R-CNN trained from scratch; A2, MRI-based non-local Mask R-CNN trained from scratch; R3, prostatectomy-based baseline Mask R-CNN trained from scratch; R4, prostatectomy-based non-local Mask R-CNN trained from scratch; R5, fine-tuned from MRI-based non-local Mask R-CNN using the 20 prostatectomy-based training patients; S1-1, self-training model using unlabelled data selected by strategy S1 at 1st iteration; S2-3, self-training model using unlabelled data selected by strategy S2 at 3rd iteration. REF, each other results compared with; --, no comparison; c1, results compared with A1; c2, results compared with A2. ***Statistically significant*** (p value <= 0.05).

| | Detection Rate (%) | P-value (c1) | P-value (c2) | DSC | P-value (c1) | P-value (c2) | 95 HD (mm) | P-value (c1) | P-value (c2) | TPR | P-value (c1) | P-value (c2) |
|---|---|---|---|---|---|---|---|---|---|---|---|---|
| A1 | 56.1% | REF | -- | 0.400 ± 0.132 | REF | -- | 8.70 ± 4.31 | REF | -- | 0.311 ± 0.140 | REF | -- |
| A2 | 63.4% | 0.252 | REF | 0.427 ± 0.140 | 0.255 | REF | 7.93 ± 3.66 | 0.294 | REF | 0.372 ± 0.175 | 0.098 | REF |
| R3 | 68.3% | 0.114 | 0.362 | 0.429 ± 0.165 | 0.259 | 0.483 | 7.65 ± 4.50 | 0.130 | 0.182 | 0.558 ± 0.268 | ***1.42E-04*** | ***0.002*** |
| R4 | 78.0% | ***0.013*** | 0.057 | 0.504 ± 0.165 | ***0.009*** | ***0.033*** | 6.66 ± 3.64 | ***0.017*** | ***0.031*** | 0.589 ± 0.222 | ***1.61E-06*** | ***9.64E-05*** |
| R5 | 75.6% | ***0.013*** | 0.065 | 0.543 ± 0.159 | ***0.001*** | ***0.003*** | 6.28 ± 3.47 | ***0.013*** | ***0.013*** | 0.625 ± 0.190 | ***1.17E-08*** | ***2.14E-06*** |
| S1-1 | 80.5% | ***0.005*** | ***0.023*** | *0.548 ± 0.165* | ***4.77E-04*** | ***0.002*** | 5.72 ± 3.17 | ***0.002*** | ***0.001*** | 0.613 ± 0.193 | ***2.80E-08*** | ***4.62E-06*** |
| S2-3 | 80.5% | ***0.002*** | ***0.012*** | 0.513 ± 0.191 | ***0.010*** | ***0.031*** | 5.81 ± 2.89 | ***0.002*** | ***0.002*** | 0.749 ± 0.190 | ***5.34E-13*** | ***1.16E-10*** |

**Table 5.** Results of models in detecting and segmenting IL, GGG > 2. R3, R4 reported with adjusted 3D aggregation thresholds with detection rate maximized. A1, MRI-based baseline Mask R-CNN trained from scratch; A2, MRI-based non-local Mask R-CNN trained from scratch; R3, prostatectomy-based baseline Mask R-CNN trained from scratch; R4, prostatectomy-based non-local Mask R-CNN trained from scratch; R5, fine-tuned from MRI-based non-local Mask R-CNN using the 20 prostatectomy-based training patients; S1-1, self-training model using unlabelled data selected by strategy S1 at 1st iteration; S2-3, self-training model using unlabelled data selected by strategy S2 at 3rd iteration. REF, each other results compared with; --, no comparison; c1, results compared with A1; c2, results compared with A2. *Statistically significant* (p value <= 0.05).

|  | Detection Rate (%) | P-value (c1) | P-value (c2) | DSC | P-value (c1) | P-value (c2) | 95 HD (mm) | P-value (c1) | P-value (c2) | TPR | P-value (c1) | P-value (c2) |
|---|---|---|---|---|---|---|---|---|---|---|---|---|
| A1 | 63.2% | REF | - | 0.425 ± 0.150 | REF | -- | 9.58 ± 4.67 | REF | -- | 0.303 ± 0.139 | REF | -- |
| A2 | 63.2% | 0.500 | REF | 0.422 ± 0.158 | 0.483 | REF | 9.69 ± 4.46 | 0.535 | REF | 0.327 ± 0.130 | 0.338 | REF |
| R3 | 89.5% | 0.063 | *0.031* | 0.469 ± 0.171 | 0.248 | 0.239 | 8.10 ± 5.26 | 0.171 | 0.112 | 0.539 ± 0.270 | *0.006* | *0.011* |
| R4 | 89.5% | *0.031* | *0.031* | 0.516 ± 0.176 | 0.086 | 0.084 | 7.51 ± 4.09 | 0.103 | 0.100 | 0.544 ± 0.217 | *0.001* | *0.003* |
| R5 | 89.5% | *0.031* | *0.031* | 0.579 ± 0.155 | *0.008* | *0.008* | 6.72 ± 3.73 | 0.063 | *0.027* | 0.593 ± 0.196 | *1.12E-04* | *2.44E-04* |
| S1-1 | 94.7% | *0.016* | *0.016* | 0.604 ± 0.135 | *0.001* | *0.001* | 6.36 ± 3.44 | *0.039* | *0.018* | 0.580 ± 0.190 | *1.28E-04* | *2.91E-04* |
| S2-3 | 84.2% | 0.063 | 0.063 | 0.631 ± 0.122 | *3.33E-04* | *3.86E-04* | 6.16 ± 3.08 | *0.043* | *0.012* | 0.746 ± 0.165 | *5.44E-08* | *1.02E-07* |

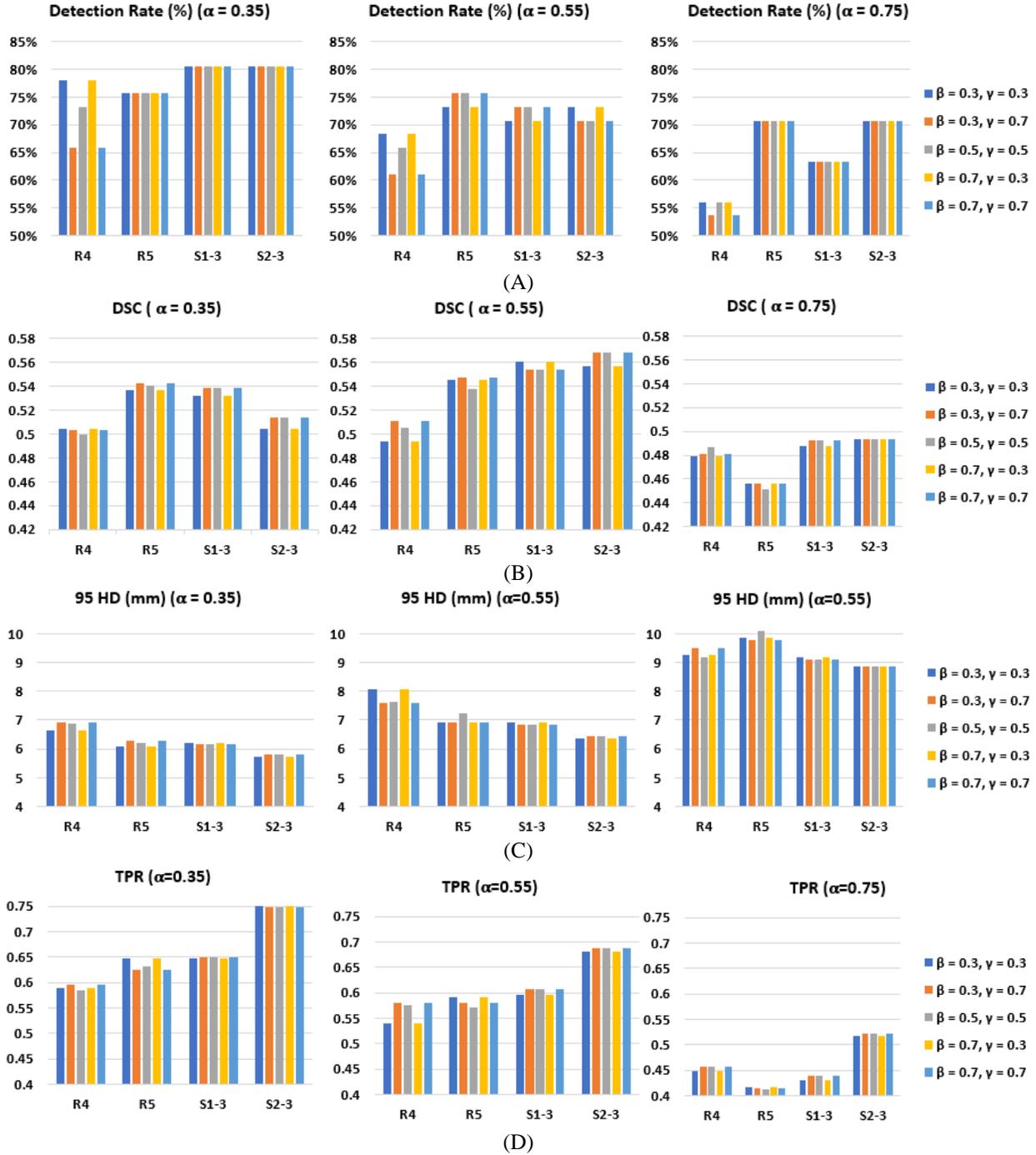

**Figure 2.** Performance of non-local Mask R-CNNs in detecting and segmenting ILs using different 3D aggregation thresholds, all GGG included. (A), detection rate (%) of top-5 predictions; (B), 3D DSC of model detected IL (true positive predictions by the model); (C) 3D 95 HD (mm) of model detected IL; (D), 3D TPR of model detected IL. R4, non-local Mask R-CNN trained from scratch using the 20 prostatectomy-based training patients; R5, MRI-based non-local Mask R-CNN (R2) fine-tuned using the 20 prostatectomy-based training patients; S1-3, self-training model using unlabelled data selected by strategy S1 at 3$^{rd}$ iteration; S2-3, self-training model using unlabelled data selected by strategy S2 at 3$^{rd}$ iteration.

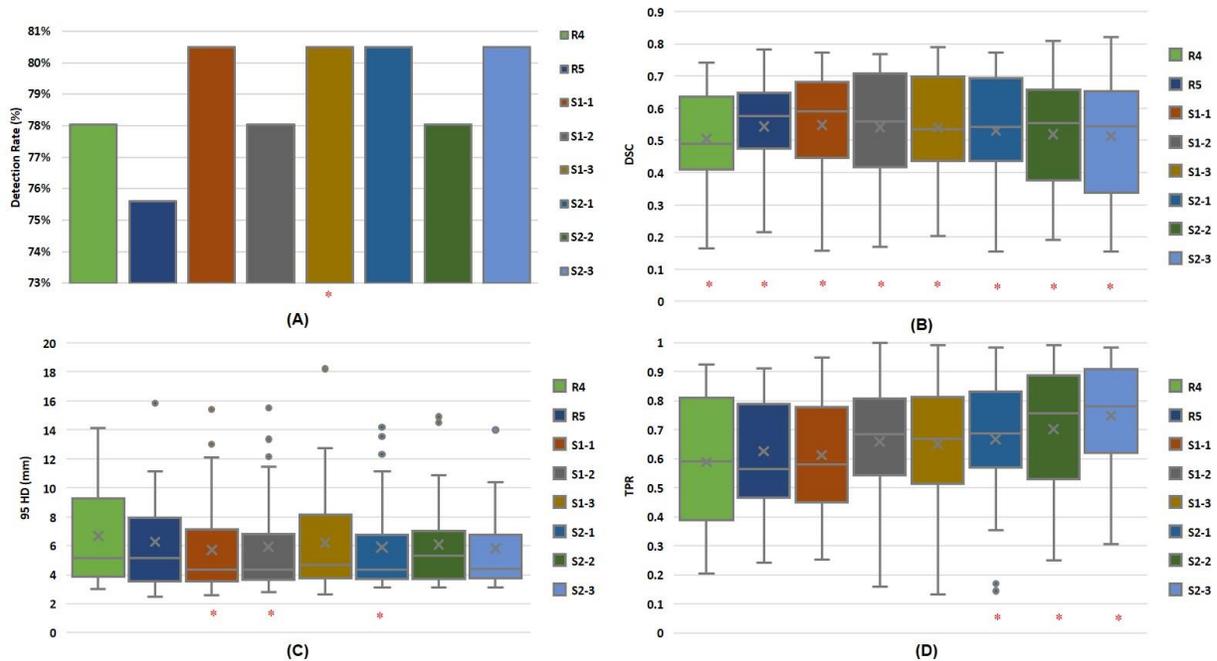

**Figure 3.** Results of prostatectomy-based non-local Mask R-CNNs in detecting and segmenting IL, all GGG. 'x' in boxplots represents for the mean value. (A), detection rate (%) of top-5 predictions; (B), 3D DSC of model detected IL (true positive predictions); (C) 3D 95 HD (mm) of model detected IL; (D), 3D TPR of model detected IL. R4, non-local Mask R-CNN trained from scratch using the 20 prostatectomy-based training patients; R5, fine-tuned from MRI-based non-local Mask R-CNN using the 20 prostatectomy-based training patients; S1-k, self-training model using unlabelled data selected by strategy S1 at kth iteration; S2-k, self-training model using unlabelled data selected by strategy S2 at kth iteration. (*) significantly improved the prostatectomy-based baseline Mask R-CNN (R3).

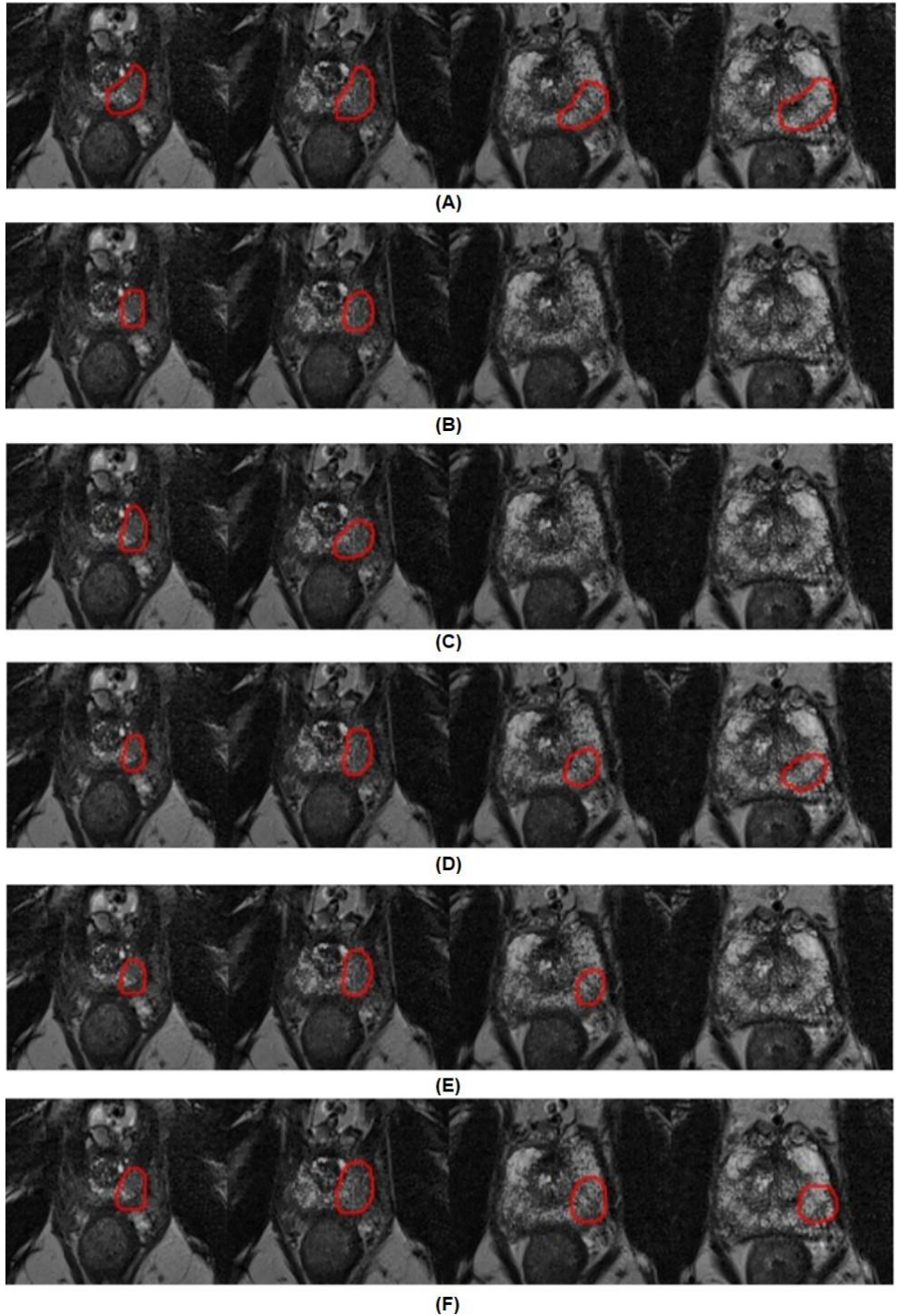

**Figure 4.** Images show examples of DL detection and segmentation of IL (GGG = 4) overlaid with T2WI. First row, prostatectomy-based ground truth delineations; second row, predictions by baseline Mask R-CNN trained from scratch (R3); third row, predictions by non-local Mask R-CNN trained from scratch (R4); forth row, predictions by fine-tuned MRI-based non-local Mask (R5); fifth row, predictions by self-training model using strategy S1 at 1$^{st}$ iteration (S1-1); sixth row, predictions by self-training using label selection strategy S2 at 3$^{rd}$ iteration (S2-3). The IL was miss by MRI-based models (A1 and A2).

## 4. Discussion

Annotations solely based on MRIs do not always align with the gold standard annotations on the histology slices. According to Merisaari et al. (10), 28 lesions annotated on the histological slices (28%, 28/99) were missed by MRI, and isotopically increasing 3D MRI-based delineations by ~10-12 mm (corresponding to Hausdorff distance) was needed to cover the entirety of prostatectomy-based delineations. This illustrates the low correlation between bp-MRI and histopathological readings. We also showed that, using prostatectomy-based delineations as ground truth with the proposed non-local Mask R-CNN significantly improved the detection and segmentation accuracy of PCa on bp-MRI. However, errors in co-registration of whole-mount prostatectomy sections with in vivo MRI can affect the results. Previous studies noted that the ex vivo shape of the prostate was considerably different from the in vivo contour depending on the time between the acquisition of the in vivo images and the prostatectomy. Tissue deformation and shrinkage occurred after surgery and toughening of tissues after fixation were contributing factors to these errors. Several semi-/fully automated methods were proposed to improve co-registration of whole-mount prostatectomy sections to in vivo MR (19-21). It is difficult to achieve 100% accurate voxel-wise classification even using ex-vivo MRI and 3D models (22). None of the automatic methods was more intelligent than human annotation. In this paper, the tissue loss, deformation and 2D distortion in sectioning were estimated visually and cognitively by a radiologist and pathologist to ensure the best ground truth. A good measure of co-registration accuracy remains a topic for future research.

Another co-registration error occurred when registering the ADC map on to T2WI and this error posed challenges in correlation of different imaging modalities when long-range dependencies were not modelled over the spatial regions. Long-range dependencies can be captured by graphical models such as conditional random fields (CRF). Artan Y et al. combined a cost-sensitive support vector machine (SVM) with CRF. T2WI, ADC and $k_{ep}$ were used, resulting in DSC of 0.56 ± 0.15 (13). However, this research was limited to the prostatic peripheral zone only. Chung AG et al. experimented with radiomics-drive CRF in 20 patients and achieved a DSC of only 0.391 (23). Other methods included: the Random Walker (RW) algorithm, where Artan Y et al. achieved a DSC of 0.57±0.21 on 16 patients (24); the Relevance Vector Machine, where Ozer S et al. achieved a DSC of 0.51 on 20 patients using the T2WI, ADC and $k_{ep}$ (25); and the Markov Random Field (MRF), where Liu X et al. achieved a DSC of 0.6222 using T2, $k_l$, IAUC, ADC, T2W, A and $k_{ep}$ (14). However, all those methods were limited to the peripheral zone, and only a small number of patients were included in their studies. The DL based models for IL segmentation have only been investigated over the past few years and are still in the early stage of development. The complexity of bp-MRI poses significant challenges with its full potential can be used to localize and segment the small ILs within the prostate gland. We identified two studies in the past of three years that conducted PCa segmentation with DL on MRI. Kohl S et al. applied an adversarial neural network for IL segmentation, resulting in a DSC of 0.41 ± 0.28; however, MRI-based delineations were used (26). Zhang G et al. proposed a bi-attention adversarial network and achieved a DSC of 0.859; however, lesion patches were cropped out before the segmentation (27). In addition, studies on 3D PCa segmentation DL models are limited, the small number of training patients, small volume size of lesions and the low axial resolution poses challenges to the training of a good 3D network. Our method eliminated the problems with 3D networks by post-aggregating 2D predictions into 3D predictions, and was an end-to-end, fully automatic method for predicting ILs within the whole prostate.

By using prostatectomy-based delineations as ground truth, the non-local Mask R-CNN (R4) described in this work has shown a significant improvement over the MRI-based baseline Mask R-CNN for detection and segmentation of IL, for all GGG. However, it took significant effort to slice and process sections in order to be able to annotate the whole mount prostatectomy specimen. It also required experienced radiologists and pathologists to work together to annotate on both MR and histologic images. Thus, it would be helpful if other techniques could further improve the performance when the amount of labelled training data is limited. Marginal improvement was observed when employing those techniques. First, comparing R5 with R4, results showed that for all GGG, fine-tuning did not improve the detection rate, and the DSC, 95 HD and TPR were improved, but not significantly. The detection rate of R4 was higher than that of R5 due to lower threshold of β (threshold of rejecting a prediction) and γ (threshold of aggregating two predictions on each slice). There were two possible reasons: 1) the mask score in R4 was low and lowering β prevented them from being rejected mistakenly; 2) by lowering γ, more 2D predictions were aggregated, and the variation was reduced by combining the binary masks and averaging the mask scores.

Next, self-training was investigated. One limitation here was that the influence of potential batch effect from different datasets was not considered, and all models were fined-tuned from the MRI-based model. Compared with R4, results showed that for all GGG, after three iterations, both selection strategies showed increased detection rate and DSC, but not significantly. For GGG > 2, the S2-3 showed significant improvement in terms of DSC (p =0.021) compared with R4, and S2-3 showed significant improvement in terms of TPR compared with R4, R5, S1, for all GGG and GGG > 2. We looked at how selected, unlabelled data impacted performance. First, by comparing the results of S1 and S2, we could infer that increasing true positives/false positives in selected unlabelled data had the corresponding impact on prediction. S2 maximized the number of true positives but also introduced more false positives compared with S1. Results showed TPR by S2 increased faster after each iteration than that by S1. The drop of DSC and increase of TPR suggested more false positives were predicted by S2 after each iteration. Second, although biopsy results and MRI-based delineations were used to select the most confident prediction out of top-5 selection, gradually decreasing DSC and increasing TPR suggested the true positives/negatives in pixels and slices also impacted the model performance. Third, compared with S1, S2 did not show a significant drop in performance, although many false positives were introduced in selected unlabelled data, suggesting that selecting unlabelled data with effort is not necessarily superior to other selection strategies. Whether semi-supervised learning outperforms a base model trained from labelled data, and whether selecting the most confident labels in self-training would be better, is controversial. Guo Y et al. found that even if all selected labels were correctly labelled, the accuracy still decreased in self-training (28). Zhou Y et al. found that not all highly confident data contributed to the performance improvement and that the trick was to select 'informative' samples, in order to reduce noise and retain overall data distribution (29). In this way, a strategy which both increases true positives and decreases false positives in selected unlabelled data is necessary, which means significantly increasing the number of patients with prostatectomy-based delineations required for overall improvement.

An additional consideration when experimenting with small volume data is whether it results in a biased or optimistic estimation of a model's performance using a simple training, validation and testing split. A 3-fold cross validation was performed to re-evaluate the model's performance, and the unlabelled data selected by S2 (iteration = 1) was also added to the training set in each fold. Table 6 shows that although testing results varied across different folds, the averaged results were all most equivalent with that of S2-1, suggesting that the predictions made were not biased with respect to the split of patients. A limitation in our work is that the prostatectomy-based patient data was acquired from a single institution. Potential variabilities introduced by different MRI scanning protocols were beyond the scope of our investigation.

Table 6. 3-fold cross validation

|  | DSC | 95 HD (mm) | TPR |
|---|---|---|---|
| fold = 1 | 0.533±0.163 | 6.26±3.06 | 0.648±0.186 |
| fold = 2 | 0.521±0.191 | 6.36±3.51 | 0.600±0.213 |
| fold = 3 | 0.551±0.159 | 6.43±3.36 | 0.597±0.202 |
| average | 0.535±0.172 | 6.35±3.31 | 0.615±0.202 |

Finally, we found low correlation between DSC and true lesion volume, suggesting the DL model did not necessarily have better segmentation accuracy in larger lesions. However, our models performed better in all the testing scenarios for the lesion group with GGG > 2, as compared to all GGG.

**5. Conclusion**
In this paper, we demonstrate state-of-art performance in automatic detection and segmentation of PCa on bp-MRIs. With prostatectomy-based delineations, the combination of non-local Mask R-CNN, fine-tuning and self-training was shown to significantly improve the detection rate, DSC, 95 HD and TPR. This model should now be further validated on a larger cohort of data with annotated MRI and whole amount prostatectomy specimens.